\documentclass[11pt]{article}
\usepackage{enumerate}
\usepackage[longnamesfirst]{natbib}
\usepackage{fancyheadings,bbm,comment}
\usepackage{multirow}
\usepackage{booktabs}
\usepackage{floatrow}
\usepackage{graphicx,subcaption}
\usepackage{arydshln,float}
\usepackage{footnote}
\usepackage{caption}
\usepackage[space]{grffile}
\usepackage{color,xcolor}
\usepackage{bm}
\makeatletter
\def\section{\@startsection{section}{1}
{\z@}{-2.5ex plus -0.5ex minus -0.1ex}{0.5ex plus 0.1ex}{\large\bfseries}}
\def\subsection{\@startsection{subsection}{2}
{\z@}{-2.25ex plus -0.3ex minus -0.2ex}{0.05ex plus 0.05ex}{\normalsize\bfseries}}
\def\subsubsection{\@startsection{subsubsection}{3}
{\z@}{-2.25ex plus -0.3ex minus -0.2ex}{0.05ex plus 0.05ex}{\normalsize\bfseries\rm}}
\makeatother
\usepackage{amsmath,amsfonts}
\usepackage{amsmath}
\usepackage{amsmath,amstext,rotating}
\usepackage{amsfonts,amssymb,graphics,xspace}

\usepackage{epsfig}

\usepackage{hypernat}
\definecolor{lust}{rgb}{0.9, 0.13, 0.13}
\definecolor{magenta(dye)}{rgb}{0.79, 0.08, 0.48}
\usepackage[
  bookmarks=true,
  bookmarksopen=true,
  breaklinks=true,
  colorlinks=true,
  linkcolor=magenta(dye),
  citecolor=magenta(dye),
]{hyperref}

\newcommand{\ba}{\begin{array}}
\newcommand{\ea}{\end{array}}
\newcommand{\bs}{\begin{align}\begin{split}\nonumber}
\newcommand{\bsnumber}{\begin{align}\begin{split}}
\newcommand{\es}{\end{split}\end{align}}

\renewcommand{\[}{\left[}

\newcommand{\bi}{\begin{itemize}}
\newcommand{\ei}{\end{itemize}}
\newcommand{\be}{\begin{enumerate}}
\newcommand{\ee}{\end{enumerate}}

\newtheorem{prop}{Proposition}

\begin{document}
\lhead{}
\rhead{\thepage}
\cfoot{}
\cfoot{\fancyplain{}}

\title{Inference on Estimators defined by Mathematical
  Programming\thanks{We thank Denis Chetverikov and Andres Santos for helpful comments.}}
\author{Yu-Wei Hsieh\thanks{Department of Economics, University of Southern California. Email: yuwei.hsieh@usc.edu}\\USC
\and Xiaoxia Shi\thanks{Department of Economics, University of Wisconsin-Madison. Email: xshi@ssc.wisc.edu}\\UW-Madison
\and Matthew Shum\thanks{Division of Humanities and Social Sciences, California Institute of Technology. Email: mshum@caltech.edu}\\Caltech}
\maketitle

\begin{abstract}
We propose an inference procedure for estimators defined by mathematical programming problems, focusing on the important special cases of linear programming (LP) and quadratic programming (QP).   In these settings, the coefficients in both the objective function and the constraints of the mathematical programming problem may be estimated from data and hence involve sampling error. Our inference approach exploits the characterization of the solutions to these programming problems by  complementarity conditions; by doing so, we can transform the problem of doing inference on the solution of a constrained optimization problem (a non-standard inference problem) into one involving inference based on a set of inequalities with pre-estimated coefficients, which is much better understood.   We evaluate the performance of our procedure in several Monte Carlo simulations and an empirical application to the classic portfolio selection problem in finance.
\end{abstract}
\vskip20pt
\baselineskip=20pt
{\footnotesize \textbf{Keywords}: Stochastic Mathematical Programming, Linear Complementarity Constraints, Moment Inequality, Duality, Sub-Vector Inference, Portfolio Selection\vskip15pt }


\newpage
In this paper, we consider the problem of inference on an estimator defined as the solution to a mathematical programming problem with pre-estimated coefficients.   Because of the pre-estimation, these coefficients contain sampling error, and hence the mathematical programming problem is stochastic.  Our focus is on the important special cases of linear programming (LP) and convex quadratic programming (QP), for which there are relevant examples in economics and finance. The difficulty with doing inference based on such estimators lies in the nondifferentiability of the estimator with respect to the data. As a result of the nondifferentiability, the estimator is not asymptotically normal, and does not allow for standard bootstrap inference (see e.g. Fang and Santos (2016)).

The core of our method lies in recognizing that the necessary and sufficient optimality conditions for LP/QP can be interpreted as inequalities with pre-estimated coefficients. Typically, these optimality conditions involve Lagrange multipliers and slackness variables for constraints, and a set of \emph{linear complementarity} (LC) conditions. Essentially, by focusing on these optimality conditions, we can transform the problem of doing inference on the solution of a constrained optimization problem (a non-standard inference problem) into one involving inference on a set of inequalities with pre-estimated coefficients, which is similar to moment inequality models and is much better understood.   Specifically, we show that the inference on the inequalities implied by the optimality conditions of LP/QP can proceed using the computationally convenient procedures from \cite{ShiShum2015}.

Estimators defined by mathematical programming have a long history in econometrics, dating back to Markowitz's  (\citeyear{Markowitz1952}) classic work on optimal portfolio selection.  More recently, \cite{ChiongGalichonShum} and \cite{ChiongHsiehShum2017} propose estimators for problems in discrete-choice analysis which also take the form of mathematical programming. Due to the absence of an inference theory, researchers often resort to bootstrap in practice; e.g., \cite{Scherer2002}. Recently, however, \cite{FangSantos2016} show that bootstrap is not valid if the solution is non-differentiable in the estimated coefficients. As the solution of mathematical programming is non-differentiable in general, our approach provides, to the best of our knowledge, the first valid inference method in the literature.  

In the next section we review the key results from the theory of linear and quadratic programming. In section 3 and 4 we provide examples and illustrate how to conduct inference respectively. In section 5 we investigate the performance of the proposed confidence set using two LP case. For the case of QP, in section 6 we estimate Markovitz's \citeyear{Markowitz1952} efficient portfolio weights and their confidence set.   As far as we are aware, our analysis of the Markowitz portfolio selection problem here represents the first instance of inference for this problem based on asymptotc approximation.

\section{Linear Programming and Quadratic Programming}
In this paper we focus on the specific cases of linear programming (LP) and quadratic programming (QP), for which our approach is easier to understand and our results are sharpest.   We will also briefly discuss more general nonlinear programming problems below.   Next we introduce the LP and QP problems in turn.

\subsection{Linear programming}
We want to estimate $\theta$ defined by the following LP:
\begin{equation}\label{lp_population}
\theta :=\textrm{argmax}\quad c'\theta\quad \textrm{s.t.}\quad A \theta\leq b
\end{equation}

where $\theta\in\Theta\subset \mathbb{R}^k$, $b$ is $m\times 1$, $c$ is $k\times 1$, and $A$ is $m\times k$. Let $A$, $b$ or $c$ be estimated from data; the sample analogs are $\hat{A}$, $\hat{b}$, and $\hat{c}$.   Then the parameter of interest is estimated by
\begin{equation}\label{lp_sample}
\hat\theta=\textrm{argmax}\quad \hat{c}'\theta\quad \textrm{s.t.}\quad \hat{A} \theta\leq \hat{b}
\end{equation}

The goal is to to derive an inference method for $\hat{\theta}$.

Our approach is to exploit the necessary and sufficient optimality conditions that characterize the solutions to linear programming problems, which follow from the duality theory of LP.   Specifically, these optimality conditions are

\begin{eqnarray}
{A}\theta&\leq& {b}\label{lcp1}\\
{A}'\lambda&=&{c} \label{lcp2}\\
\lambda&\geq& 0\\
{c}'\theta&=& b'\lambda \label{lcp4}
\end{eqnarray}

where $\lambda\in\mathbb{R}^m_{+}$ (See \cite{Mangasarian1969}, p. 18.).\footnote{In optimization theory, these conditions are the basis for the primal-dual interior point method for solving LP.} Equation (\ref{lcp1}) and (\ref{lcp2}) express, respectively, primal and dual feasibility, where $\lambda$ is interpreted as the $m\times 1$ vector of Lagrange multipliers on the inequalities (\ref{lcp1}).\footnote{Recall the dual LP problem corresponding to (\ref{lp_population}) is $\min_{\lambda\geq 0} b'\lambda$ subject to $A'\lambda=c$.} The final equation (\ref{lcp4}) is a complementarity condition, analogous to the complementarity slackness condition in Karush-Kuhn-Tucker (KKT) conditions.\footnote{Combining (\ref{lcp2}) and (\ref{lcp4}), we obtain $\lambda'(b-A\theta)=0$, which is the usual complementary slackness condition for this problem.}  These considerations yield the following key proposition.

\begin{prop} (LP Inference).
Inference on $\hat\theta$ defined as the solution to the LP problem (\ref{lp_sample}) is equivalent to inference on $\hat\theta$ satisfying the inequalities (\ref{lcp1})-(\ref{lcp4}) evaluated at the estimated quantities $\hat{A}, \hat{b}, \hat{c}$.
\label{prop:lp}\end{prop}

Accordingly, our inference procedure exploits the fact that the optimality conditions (\ref{lcp1})-(\ref{lcp4}) are just a set of linear equalities and inequalities in the unknowns $\theta$ and $\lambda$.  Therefore, inference on these conditions falls into the special class of inequality models considered in \cite{ShiShum2015}, for which computationally attractive procedures (not involving time-consuming bootstrap steps) are available for constructing joint confidence sets for $(\theta,\lambda)$, and projected confidence sets for $\theta$.

More broadly, by exploiting the optimality conditions (\ref{lcp1}-\ref{lcp4}), we can transform the problem of inference on a LP problem, which is difficult, to inference on parameters defined by a set of linear inequalities, which is a relatively straightforward exercise with an existing literature.\footnote{Indeed, characterizing the solution to a constrained optimization problem via the optimality conditions (\ref{lcp1}-\ref{lcp4}) is analogous to characterizing the solution to an unconstrained optimization problem using the first-order conditions, which underlies the usual approach for doing inference with M-estimators.}

\subsection{Quadratic Programming}
A second class of problems covered by our method is the convex Quadratic Programming (QP) case.\footnote{\cite{wolak1987exact} also exploits the duality theory for nonlinear programming in deriving test statistics for nonlienar parameter constraints in the linear regression model.}

\begin{align}
\begin{array}{ll}\label{QP}
\min & c'\theta+\frac{1}{2}\theta'Q\theta\\
\textrm{s.t.}& A\theta\geq b,\\
\end{array}
\end{align}

where $Q$ is positive semi-definite. In this case, the KKT conditions are both necessarily and sufficient (see \cite{CottlePangStone1992}, p. 4). These conditions are, first,  primal feasibility:

\begin{equation}\label{QP_primal}
A\theta-b-s=0;
\end{equation}

second, dual feasibility 
\begin{equation}\label{QP_dual}
A'\lambda-c-Q\theta=0;
\end{equation}

and finally, the complementarity conditions
\begin{align*}
\lambda's&=0 \\
\lambda&\geq 0\\
s & \geq 0.
\end{align*}

Because both $\lambda_i$ and $s_i$ are non-negative, it follows that $\lambda' s=0$ is equivalent to $\lambda_is_i=0\quad\forall i$. Using shorthand from the optimization literature,  we write them collectively as

\begin{equation}\label{lcqp}
0\leq \lambda_i\perp s_i\geq 0.
\end{equation}

For inference, we consider the case where the coefficients in the QP problem, $(A,b,c,Q)$ are estimated and thus contain sampling error.
Analogously to Proposition~\ref{prop:lp}, we have:

\begin{prop}\label{prop:qp}
  (QP Inference).
Inference on $\hat\theta$ defined as the solution to the QP problem (\ref{QP}) is equivalent to inference on $\hat\theta$ satisfying the inequalities (\ref{QP_primal}),(\ref{QP_dual}),(\ref{lcqp}) evaluated at the estimated quantities $\hat{A}, \hat{b}, \hat{c}, \hat{Q}$.
  \end{prop}
As in the LP case, the QP optimality conditions in Eqs. (\ref{QP_primal}), (\ref{QP_+dual}), and (\ref{lcqp}) are linear inequalities with pre-estimated coefficients in the parameters, and the simpler inference methods in \cite{ShiShum2015} are applicable.  

\subsection{Related literature}
As far as we are aware, we are among the first to set forth inference theory for a quantity ($\hat\theta$) which is a solution to a ``noisy'' LP or QP problem, where there is some noise to the sample or estimation error in the constraints.\footnote{In the engineering literature, LP with noisy model parameters is also extensively studied under the umbrella of \emph{robust linear programming}. The goal in robust LP is to obtain a \emph{single} solution $\theta$ which remains ``optimal'' in the presence of estimation error. In constrast, our goal is to solve the statistical inference problem of obtaining a \emph{set of solutions} -- the confidence set -- that can include the true solution with pre-specified probability.} Our approach is to exploit the optimality conditions (\ref{lcp1}-\ref{lcp4}) to show that doing inference on $\hat{\theta}$ (defined in \ref{lp_sample}) is equivalent, as testing the inequality constraints (\ref{lcp1}-\ref{lcp4}).   That is, the confidence set for $\theta$ defined as the optimizer of (\ref{lp_sample}) is equivalent to the confidence set for $\theta$ we get by ``inverting'' the test of the inequalities formed from the optimality conditions of the underlying mathematical programming.

Our paper is related to work by Wolak (\citeyear{wolak1987exact}, \citeyear{wolak1989local}, \citeyear{wolak1989testing}) on testing (in)equality constraints on parameters in linear and nonlinear econometric models.  The duality in mathematical programming problems plays an important role in Wolak's analysis, as it does in ours; however, he considers the case where the constraints are deterministic, while we focus on the case where the coefficients in the constraints are subject to sampling error.   \cite{GuggenbergerHahnKim08} derive specification tests for moment inequality models by exploiting dual formulations of the constraints, but  not in a mathematical  programming context.

\cite{KaidoSantos2014} on the other hand transform the moment inequality model into a convex programming problem. Their main focus is efficient estimation of moment inequality models. \cite{FreybergerHorowitz2015} consider inference for optimized linear programming objective functions ($\max_{\theta} c'\theta$) rather than the optimizing parameter $\textrm{argmax}_{\theta} c'\theta$, which is the focus here.  Similarly, \cite{KaidoMolinariStoye2016} consider inference on the optimized criterion $\max_{\theta} c'\theta$ subject to nonlinear moment inequalities.   Indeed, our approach may also work if the object of interest is the optimized criterion; in that case, we can introduce another parameter $\rho=c'\theta$ and consider the LP problem of $\max_{\rho, \theta} \rho$ subject to $A\theta\leq b$ and $\rho=c'\theta$.

\section{Examples}
Here we present several examples.  The first four examples are linear programming estimators; the final one is a quadratic programming estimator.

{\bfseries Example 1: Intersection bounds.} We want to estimate $\theta=\max \left\{ \mathbb{E}X_1, \mathbb{E}X_2\right\}$, which is a very simple example of intersection bounds.\footnote{See, eg., \cite{chernozhukov2013intersection}, \cite{FangSantos2016}.}  This can be written as the solution (argmax) of the linear programming problem
  $$
  \min_{\theta} \theta\quad s.t. \quad \theta\geq \mathbb{E}X_1, \theta\geq \mathbb{E}X_2.\label{example:interbnd}
  $$

{\bfseries Example 2: Market share prediction in semiparametric discrete choice models \cite{ChiongHsiehShum2017}.} We wish to predict market shares in a semiparametric multinomial choice demand model.   We observe market shares and covariates across $M$ markets: $\left\{ \mathbf{s}_m, \mathbf{X}_m\right\}_{m=1}^M$.  Assume we use \cite{ShiShumSong2017} to estimate parameters in utility: $U^k_m=\beta X^k_m$.   Now we have a counterfactual market $M+1$ with covariates $\mathbf{X}_{M+1}$.   The market shares $\mathbf{s}_{M+1}$ are not point identified, but must satisfy the cyclic monotonicity conditions taken across markets $m=1,2,\ldots,M, M+1$.   Formally we estimate
  $$
\max_{\mathbf{s}_{M+1}} s^k_{M+1}\quad s.t. \quad CM(\mathbf{s}_{M+1}; \hat\beta, \left\{ \mathbf{s}_m, \mathbf{X}_m\right\}_{m=1}^M, \mathbf{X}_{M+1}).
$$
CM denotes the linear inequalities arising from cyclic monotonicity.   For instance, if we consider only length-2 cycles, then they are, for all $m\in \left\{1,2,\ldots,M\right\}$:
$$
(\mathbf{s}_{m}-\mathbf{s}_{M+1}) (\mathbf{X}_{M+1}'-\mathbf{X}_{m}')\hat\beta\leq 0.
$$

We may be interested in other quantities.  For instance, for a multiproduct firm which produces goods (say) 1,2,3, the highest counterfactual revenue is
$$
\max_{\mathbf{s}_{M+1}} \sum_{k=1,2,3} {p}^k_{M+1} {s}^k_{M+1}\quad s.t. \quad CM(\mathbf{s}_{M+1}; \hat\beta, \left\{ \mathbf{s}_m, \mathbf{X}_m\right\}_{m=1}^M, \mathbf{X}_{M+1})
$$
and the market shares of (say) good 2 among the set of revenue-maximizing market shares would be the argmax of this problem.

{\bfseries Example 3: bounds on nonparametric regression function subject to shape restrictions.}  Following \cite{FreybergerHorowitz2015}, consider a nonparametric regression model $Y=g(X)+U$ with $\mathbb{E}[U|W=w]=0\ \forall w$; here $Y$ is an outcome of interest, $X$ is a possibly endogenous regressor and $W$ is an instrument (and both $X$ and $W$ are finite-valued).   We may wish to derive bounds on values of the finite-valued unknown function $g$ which maximize a linear functional $c'g$ subject to shape restrictions:
$$
\text{argmax}_g c'g \quad\text{s.t.} \quad \Pi'g=m; \ Sg\leq 0.
$$

{\bfseries Example 4: Nonparametric utility estimation in discrete-choice demand models.} \cite{ChiongGalichonShum} show that the utility indices can be non-parametrically recovered from the market shares/choice probabilities, in the additive random utility model with known distribution of utility shocks. Suppose individual $i$ obtains utility $\alpha_j+\epsilon_{ij}$ by choosing alternative $j$, where the joint distribution $\epsilon_{i\cdot}\sim G$. The researcher also has the estimated market shares $p_j$, and she can simulate $N$ random draws of $\epsilon_{i\cdot}$ from $G$. They show that $\alpha_j=-v_j$, where $v_j$ solves the following linear programming problem:

\begin{align}
\begin{array}{ll}
\min_{(u,v)}&\sum_{i=1}^N u_i + N\sum_{j=1}^{J} p_{j}v_{j}\\[10pt]
\textrm{s.t.}& u_{i}+v_{j}\geq \epsilon_{ij}\forall i,j
\end{array}
\end{align}

In this case, conditional on the simulator $\epsilon$, there is no uncertainty in $(A,b)$. Part of the objective coefficient $c$ is subject to the estimation error.

{\bfseries Example 5: Optimal portfolio selection.} Our final example is one involving quadratic programming, and will be our empirical example below.   One of the most famous QP problem in economics is the modern portfolio theory of \cite{Markowitz1952}. Suppose there are $k$ assets, with expected return $R$, and covariance matrix for the return on these assets $Q$. In practice these two quantities are estimated from return data. $\theta$ is the portfolio weight vector such that $\sum_{i=1}^{k}\theta_i=1$.\footnote{Negative weight means short position.} Clearly, $\theta'Q \theta$ is the variance of portfolio return and $R'\theta$ is the expected return on the portfolio. Given a targeted expected return $\mu$, \cite{Markowitz1952} considers the minimum risk, long-only portfolio by solving the following QP problem:

\begin{align}\label{QP_portolio}
\begin{array}{ll}
\min &\theta'Q\theta\\[10pt]
\textrm{s.t.}& R'\theta=\mu\\[10pt]
&\bm{1}'\theta=1\\[10pt]
&\theta\geq0
\end{array}
\end{align}

\section{Inference on parameter vector $\theta$}
In this section we detail the inference procedure for LP with all-inequality constraints.\footnote{The case of QP is similar and is discussed in Section 5 below.} In order to apply the computationally simple procedure of \cite{ShiShum2015}, we first introduce the $m\times 1$ vector ${s}$ of nonnegative slackness parameters.\footnote{Recently, \cite{Chenetal2016} also introduce slackness parameters to transform moment inequalities to equalities in their MCMC-based estimator for partially identified models.}  Then we can rewrite the primal-dual feasibility and linear complementarity conditions (\ref{lcp1}-\ref{lcp4}) as:

\begin{align}\label{newlcp1}
{A}\theta + s-b&= 0\\
{A}'\lambda-c&=0 \label{newlcp2}\\
\lambda's&=0 \label{newlcp4}\\
\lambda&\geq 0\\
s & \geq 0
\end{align}

In this version of primal-dual formulation, the components of the model estimated with sampling error -- $(A,b,c)$ -- enter only the equalities (\ref{newlcp1} - \ref{newlcp2}),  while the LC condition (\ref{newlcp4}) serves as the nonlinear constraints on parameters. Thus this falls into the framework of \cite{ShiShum2015}. For equality constraints, there is no complementary slackness and non-negative constraint for $\lambda$. Eq. (\ref{newlcp1}) and Eq. (\ref{newlcp2}) are modified as
\begin{align}\label{newlcp1}
{A}\theta -b&= 0\\
{A}'\lambda-c&=0
\end{align}

Let $g(A, b, c,  \theta,\lambda,s)=\left(\begin{array}{c}A\theta+s-b\\ A'\lambda-c\end{array}\right)$. For any $m\times k$ matrix $W$, let $\textup{vec}({W}) = (W_{\cdot,1}',\dots,W_{\cdot,k}')'$ where $W_{\cdot,j}$ is the $j$th column of $W$. Suppose that $X$ is another matrix. Let the Kronecker product of $W$ and $X$ be denoted $W\otimes X$, i.e.

$$
W\otimes X = \begin{pmatrix}w_{11}X,\dots,w_{1k}X\\\vdots,\ddots,\vdots\\ w_{m1}X,\dots,w_{mk}X\end{pmatrix}.
$$

\medskip
Using the newly introduced notation, we can write $g(A, b,c,\theta,\lambda,s)$ as

\begin{align}\label{eq:moment_condition}
g(A, b,c,\theta,\lambda,s) &= \begin{pmatrix}(\theta'\otimes I_m)\textup{vec}(A) +s - b\\
(I_k\otimes \lambda')\textup{vec}(A) - c\end{pmatrix}\nonumber\\
&=\begin{pmatrix}\theta'\otimes I_m&-I_m&0_{m\times k}\\
I_k\otimes \lambda'&0_{k\times m}&I_k
\end{pmatrix} \begin{pmatrix}\textup{vec}(A)\\b\\c\end{pmatrix} +\begin{pmatrix}s\\0_{k\times 1}\end{pmatrix}.
\end{align}

\bigskip
Let $G(\theta,\lambda,s) = \begin{pmatrix}\theta'\otimes I_m&-I_m&0_{m\times k}\\
I_k\otimes \lambda'&0_{k\times m}&I_k\end{pmatrix}$.
Suppose that $A, b, c$ are estimated by $\widehat{A}, \hat{b}, \hat{c}$.  Assume that
\begin{equation}\label{eq:asym_cov}
\sqrt{n}\begin{pmatrix}\textup{vec}(\widehat{A})-\textup{vec}({A})\\\hat{b}-b\\\hat{c}-c\end{pmatrix}\to _d N(0, V).
\end{equation}

Let $V$ be estimated by $\widehat{V}$. Let
\begin{equation}
\widehat{Q}_n(\theta,\lambda,s) = g(\widehat{A},\hat{b},\hat{c},\theta,\lambda,s)'(G(\theta,\lambda,s)\widehat{V}G(\theta,\lambda,s)')^{-1}g(\widehat{A},\hat{b},\hat{c},\theta,\lambda,s)
\end{equation}

Following \cite{ShiShum2015}, the confidence set of confidence level $1-\alpha$ can be constructed as
\begin{align}\label{eq:CS_SS}
CS^{SS}_n(1-\alpha) = \{\theta\in \Theta:\min_{ \lambda\geq 0, s\geq 0:\lambda's=0} n\widehat{Q}_n(\theta,\lambda,s) \leq \chi^2_{m+k}(1-\alpha)\},
\end{align}

Computing the profile test statistics itself only involves a GMM objective function of linear moments, subject to LC constraints. This falls into the class of ``Mathematical Programming with Complementarity Constraints (MPCC)'' problems which are well-understood computationally.\footnote{MPCC problems can be easily specified in the KNITRO interface for MATLAB; see \url{https://www.artelys.com/tools/knitro_doc/2_userGuide/complementarity.html}.} Therefore our method is user-friendly and is not computationally demanding.\footnote{See \cite{DongHsiehShum2017} for additional applications of MPCC in general moment inequality models.}

In practice, it is usually convenient to report the upper and lower bound of the confidence set of each parameter. For example, for the parameter $\theta_j$, one can report the confidence interval $[\underline{\theta}_{j}(1-\alpha), \overline{\theta}_{j}(1-\alpha)]$, which can be obtained by solving the following problems:

\begin{align}
\underline{\theta}_j(1-\alpha)& = \inf \theta_j,\quad \text{st. }{\theta\in \Theta: \theta\geq 0, \min_{\lambda\geq 0,s\geq 0:\lambda's=0 }n\widehat{Q}_{n}(\theta,\lambda,s)\leq \chi^2_{m+k}(1-\alpha)}; \nonumber\\
\overline{\theta}_j(1-\alpha) &= \sup \theta_j,\quad \text{st. }{\theta\in \Theta: \theta\geq 0, \min_{\lambda\geq 0,s\geq 0:\lambda' s=0 }n\widehat{Q}_{n}(\theta,\lambda,s)\leq \chi^2_{m+k}(1-\alpha)}.\label{proj-ci}
\end{align}

\section{Monte Carlo Simulations}
Next we consider two simulation examples, for the LP case.
\subsection{Simulation 1: Intersection bounds.}
We first consider a simple intersection bounds problem mentioned in example 1. This can be reformulated as a LP with scalar parameter ($k=1$) and $m=2$ constraints.   Thus, for compatibility as stated in Eq. (\ref{lp_population}), we have  $A=[-1 -1]'$, $c=-1$, and $b=[-\mathbb{E}X_1, -\mathbb{E}X_2]'$.   For this example we introduce two slackness parameters $s=[s_1, s_2]'$ and also two Lagrange multiplier $\lambda=[\lambda_1, \lambda_2]$. In this case, we have the following sample moment conditions:

\begin{equation}
\begin{array}{l}
\frac{1}{N}\sum_iX_{1i}-\theta+s_1=0\\[8pt]
\frac{1}{N}\sum_iX_{2i}-\theta+s_2=0\\[8pt]
\lambda_1+\lambda_2-1=0\\[8pt]
0\leq \lambda_1\perp s_1\geq 0\\[8pt]
0\leq \lambda_2\perp s_2\geq 0
\end{array}
\end{equation}

Because here we have non-stochastic $A$ and $b$, only the first two equations (primal feasibility) are treated as moment conditions defined in (\ref{eq:moment_condition}). The dual feasibility and LC conditions are treated as constraints when computing the profile test statistics in (\ref{eq:CS_SS}):

$$
\min_{\lambda_1+\lambda_2=1, \lambda_i\geq 0, s_i\geq 0:\lambda_is_i=0} n\widehat{Q}_n(\theta,\lambda,s)
$$

The asymptotic variance $\hat{V}$ defined in (\ref{eq:asym_cov}) is the sample covariance matrix of $(X_{1i},X_{2i})$, while and Jacobian $G(\theta,\lambda,s)$ is the identity matrix. We report the coverage probability under different combinations of DGP and sample size in Table \ref{tab:1}. From Table \ref{tab:1} we can see that the empirical coverage rate is slightly greater than the pre-specified confidence level. When the sample size increases, our confidence set also becomes more conservative. This over-conservative finding is consistent with the simulation results in \cite{ShiShum2015}, which is a general issue in sub-vector inference in inequality models inclulding moment inequality models.

\subsection{Simulation 2}
Consider the following population LP problem (\ref{lp_population}) with
\begin{align}
A=
\begin{pmatrix}
1&2\\
1&-1\\
\end{pmatrix}, b=\begin{pmatrix}
4\\
1\\
\end{pmatrix}, c=\begin{pmatrix}
3\\
2\\
\end{pmatrix}.
\end{align}

We further impose the solution $\theta$ is non-negative. We use these numbers as population means and generate normal random numbers with variance 1, and then compute the corresponding sample means $\hat{A},\hat{b},\hat{c}$. The solution of the population LP problem is $\theta=(2,1)$. We report the empirical coverage rate of the confidence set (\ref{eq:CS_SS}) in Table \ref{tab:2}. Similar to the previous experiment, the confidence set is over-conservative.

\section{Empirical Illustration: Portfolio Selection}
In this section we illustrate how to compute the confidence set for \cite{Markowitz1952}'s efficient portfolio weights. In the portfolio selection problem (\ref{QP_portolio}), there are two primal feasibility conditions

\begin{align}
\begin{array}{ll}
R'\theta-\mu&= 0\\[10pt]
\bm{1}'\theta-1&=0\\
\end{array}
\end{align}

and $k$ dual feasibility conditions:

\begin{align}
\begin{array}{l}
\bm{\lambda}_{\theta}+\lambda_{R}R+\lambda_{F}\bm{1}-Q\theta=\bm{0},
\end{array}
\end{align}

where $\bm{\lambda}_{\theta}$ is the vector of Lagrange multipliers of the non-negative constraints, and $(\lambda_{R},\lambda_{F}$ are respectively the Lagrange multipliers of the equality constraints of targeted return and feasible portfolio weights. There are $k$ linear complementarity conditions: $0\leq\bm{\lambda}_{\theta}\perp \theta\geq0$. Here $\theta$ are both the \emph{decision variables} as well as the \emph{slackness variables}. Because the portfolio weight constraint $\bm{1}'\theta-1=0$ does not involve estimated moments, we exclude it from the moment conditions. The rest of primal-dual conditions can then be expressed in terms of the following moment conditions $g(\hat{Q},\hat{R},\mu;\theta,\lambda)=$

$$
\begin{pmatrix}
\hat{R}'&\bm{0}'&0&0\\
-\hat{Q}&I_{k\times k}&\hat{R}&\bm{1}\\
\end{pmatrix}\cdot\begin{pmatrix}
\theta\\
\bm{\lambda}_{\theta}\\
\lambda_R\\
\lambda_F\\
\end{pmatrix}-\begin{pmatrix}
\mu\\
\bm{0}
\end{pmatrix}
$$
$$
=\begin{pmatrix}
\bm{0}_{1\times k^2}&\theta^{'}\\
-\theta^{'}\otimes I_{k\times k}&\lambda_{R}I_{k\times k}
\end{pmatrix}\cdot\begin{pmatrix}
\mathrm{vec}(\hat{Q})\\
\hat{R}
\end{pmatrix}+\begin{pmatrix}
-\mu\\
\bm{\lambda}_{\theta}+\lambda_{F}\bm{1}
\end{pmatrix}
$$

The portfolio feasibility constraint $\bm{1}'\theta-1=0$ is instead treated as a parameter constraint when computing the test statistics:

$$
CS^{SS}_n(1-\alpha) = \{\theta\in \Theta:\min_{ \bm{1}'\theta-1=0, 0\leq\bm{\lambda}_{\theta}\perp \theta\geq0} n\widehat{Q}_n(\theta,\lambda) \leq \chi^2_{1+k}(1-\alpha)\}.
$$

\begin{figure}
\begin{center}
\includegraphics[width=1.1\textwidth]{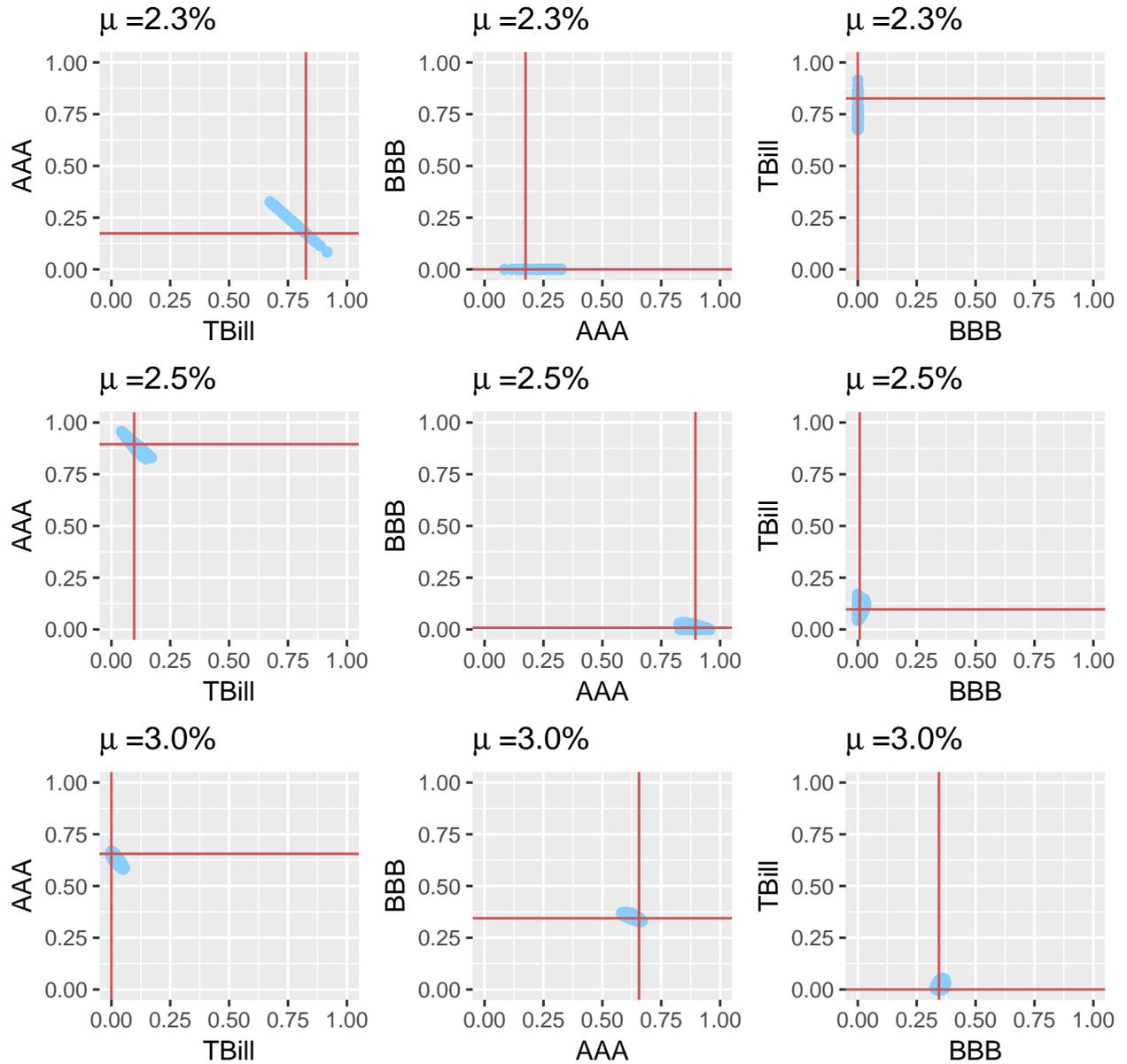}
\end{center}
\caption{90\% Confidence Set of Efficient Portfolio Weights under Different Target Return $\mu$} \label{Fig:CS}
\footnotesize{\emph{Note:} The solution of portfolio selection (\ref{QP_portolio}) based on the estimated $(\hat{R},\hat{Q})$ is located by two red lines.}
\end{figure}

We consider portfolio selection over three fixed income securities: 10-year Treasury Bill, AAA corporate bond and BBB corporate bond. We calculate the annualized return and covariance matrix using the daily data from 2010-01-04 to 2017-07-31\footnote{We use 10-Year Treasury Constant Maturity Rate, and BofA Merrill Lynch US Corporate AAA and BBB Effective Yield, downloaded from Federal Reserve Bank of St. Louis, \url{https://fred.stlouisfed.org/}.}

\begin{align}
\hat{R}=
\begin{pmatrix}
\text{T-Bill}\\
\text{AAA}\\
\text{BBB}\\
\end{pmatrix}=\begin{pmatrix}
2.2550\\
2.5137\\
3.9256\\
\end{pmatrix}, \hat{Q}=\begin{pmatrix}
0.5976&\cdot&\cdot\\
0.2336&0.2674&\cdot\\
0.2758&0.2285&0.4488\\
\end{pmatrix},
\end{align}

where the unit of measurement is percentage. We use grid search to depict the confidence set under different target return $\mu$. The results are reported in Figure \ref{Fig:CS}. Although our confidence set may be over-conservative, in this case it does yields tight estimation. The upper panel of Figure \ref{Fig:CS} suggests that, when the target return is low ($\mu=2.3\%$), the confidence set is a set of linear combinations among T-Bill and AAA Corporate bonds; one never hold a positive position on the BBB Corporate bonds. On the other hand, at $\mu=3\%$, the confidence set depicted in the lower panel of Figure \ref{Fig:CS} has a more standard ``elliptical'' shape.

The confidence set is potentially useful for making investment decisions. When new data becomes available, the portfolio weight based on the newly calculated $(\hat{R},\hat{Q})$ can be substantially different from the old one. It thus raises the question whether one should adjust the portfolio or view the difference as within tolerance given given the estimation error due to finite sample noise. Our confidence set can be used to addressed this issue; one simply test whether the previous $\theta$ belongs to the confidence set based on the latest data. To the best of our knowledge, the distributional theory for efficient portfolio is only available under the assumption of allowing for short position and normality of the return data; see \cite{JobsonKorkie1980}. For the general case one usually relies on bootstrap test (\cite{Scherer2002}). However, bootstrap is not valid in light of \cite{FangSantos2016}. Our method is valid  in the general case as it can accommodate different types of constraints commonly encountered in practice, as well as weaker requirement for the underlying DGP.

\section{Conclusion}
We propose an inference procedure for estimators defined as optimizers of stochastic versions linear and quadratic programming problems with pre-estimated coefficients in the objective function or constraints. The Karush-Kuhn-Tucker conditions which characterize the optimum are re-interpreted as inequalities with pre-estimated coefficients, inference on which can be carried out as special cases of \cite{ShiShum2015}.   We provide an empirical application to the portfolio selection problem in finance; as far as we are ware, this represents the first instance of inference for this classic problem based on asymptotc approximation.

More broadly, since KKT conditions naturally arise from convex programming problems, our inference approach might also work in those more general contexts. When the resulting inequalities are moment inequalities, one can use the well-established methods in the moment inequality literature (e.g. \cite{AndrewsSoares2010}, and \cite{AndrewsBarwick2012}, among others) to construct joint confidence sets for $(\theta,s,\lambda)$ and then obtain the marginal confidence set for $\theta$ as projection of the joint confidence sets. For methods that focuses on marginal confidence sets for $\theta$ (which usually yield tighter inference than the simple projection method above), one could use more elaborate methods like \cite{KaidoMolinariStoye2016} and \cite{BugniCanayShi2017} in case of moment inequality models, and \cite{Kim2017} in other cases.

\bigskip
\linespread{1}\small
\bibliographystyle{biblioStyleEconometrica}
\bibliography{refLP}

\newpage
\section*{Tables}
\begin{table}[htbp]
  \centering
  \caption{Simulation 1: Empirical Coverage Rate at 95\% Confidence Level. \cite{ShiShum2015} Confidence Set}
    \begin{tabular}{lrrr}
    \toprule
     &\multicolumn{3}{c}{sample size}\\
          & \multicolumn{1}{l}{n=100} & \multicolumn{1}{l}{n=200} & \multicolumn{1}{l}{n=500} \\
   $\theta=\max\{\mathbb{E}X_1,\mathbb{E}X_2\}$\footnote{$(X_1,X_2)\sim_d N(\mu, \Sigma)$. 1000 Monte Carlo Repetitions.}&&&\\
   \cmidrule{2-4} design 1\footnote{$\mu=(5,3), (\sigma^2_1,\sigma^2_2)=(1,1), \sigma_{12}=0$}& 0.983	&0.983&	0.989 \\
     design 2\footnote{$\mu=(5,3), (\sigma^2_1,\sigma^2_2)=(3,1), \sigma_{12}=0$} & 0.983&0.987&0.993\\
     design 3\footnote{$\mu=(5,3), (\sigma^2_1,\sigma^2_2)=(3,1), \sigma_{12}=1.5$}& 0.984&0.986&0.984\\
    \bottomrule
    \end{tabular}
  \label{tab:1}
\end{table}

\begin{table}[htbp]
  \centering
  \caption{Simulation 2: Empirical Coverage Rate at 95\% Confidence Level. \cite{ShiShum2015} Confidence Set}
    \begin{tabular}{lrrr}
    \toprule
     &\multicolumn{3}{c}{sample size}\\
     & \multicolumn{1}{l}{n=100} & \multicolumn{1}{l}{n=200} & \multicolumn{1}{l}{n=500} \\

     true parameter $=(2,1)$\footnote{1000 Monte Carlo Repetitions}& 0.986 & 0.99 & 0.996 \\
    \bottomrule
    \end{tabular}
  \label{tab:2}
\end{table}

\end{document}